\documentclass[iop]{emulateapj-rtx4}
\usepackage{bm}
\usepackage{threeparttable}
\usepackage{CJK}
\def\msun{M_\odot}
\def\mbh{M_{\rm{BH}}}

\newcommand\bI{{{\sf\boldmath I}}}

\def\<{\,\langle\langle}
\def\>{\,\rangle\rangle}

\begin{document}
\begin{CJK*}{UTF8}{gbsn}

\shortauthors{Y.-F. Jiang et al.}
\author{Yan-Fei Jiang(姜燕飞)\altaffilmark{1,2}\footnote{Einstein Fellow}, James M. Stone\altaffilmark{2} \& Shane W. Davis\altaffilmark{3}}
\affil{$^1$Harvard-Smithsonian Center for Astrophysics, 60 Garden Street, Cambridge, MA 02138, USA} 
\affil{$^2$Department of Astrophysical Sciences, Princeton
University, Princeton, NJ 08544, USA} 
\affil{$^3$Canadian Institute for Theoretical Astrophysics. Toronto, ON M5S3H4, Canada}

\title{Radiation Magneto-hydrodynamic Simulations of the Formation of Hot Accretion Disk Coronae}

\begin{abstract}
A new mechanism to form a magnetic pressure supported, high temperature
corona above the photosphere of an accretion disk is explored
using three dimensional radiation magneto-hydrodynamic (MHD)
simulations. The thermal properties of the disk are calculated
self-consistently by balancing radiative cooling through the surfaces
of the disk with heating due to dissipation of turbulence driven
by magneto-rotational instability (MRI).  As has been noted in
previous work, we find the dissipation rate per unit mass increases
dramatically with height above the mid-plane, in stark contrast to
the $\alpha$-disk model which assumes this quantity is a constant.
Thus, we find that in simulations with a low surface density (and
therefore a shallow photosphere), the fraction of energy dissipated
above the photosphere is significant (about 3.4\% in our
lowest surface density model), and this fraction increases as surface density
decreases.  When a significant fraction of the accretion energy is
dissipated in the optically thin photosphere, the gas temperature
increases substantially and a high temperature, magnetic pressure
supported corona is formed. The volume-averaged temperature in the disk
corona is more than $10$ times larger than at the disk mid-plane.
Moreover, gas temperature in the corona is strongly anti-correlated
with gas density, which implies the corona formed by MRI turbulence
is patchy.  This mechanism to form an accretion disk corona may
help explain the observed relation between the spectral index and
luminosity from AGNs, and the soft X-ray excess from some AGNs.  It
may also be relevant to spectral state changes in X-ray binaries.

\end{abstract}


\maketitle

\section{Introduction}
The standard $\alpha$-disk model \citep[][]{ShakuraSunyaev1973} is
usually used to account for the thermal emission from both AGN
\citep[e.g.,][]{Krolik1999} and X-ray binaries
\citep[e.g.,][]{Doneetal2007}. However, a hard X-ray tail, or a
soft X-ray component with energies much larger than that expected
from thermal emission from the disk are often observed from both
AGNs \citep[][]{Elvisetal1978} and X-ray binaries
\citep[][]{RemillardMcClintock2006,Doneetal2007}. During a state
transition in X-ray binaries, the hard X-ray tail may actually
become stronger than the thermal emission from the accretion disk
\citep[][]{RemillardMcClintock2006}.

Hard X-rays from black hole accretion flows are usually believed
to be formed via inverse Compton scattering of seed photons from
the accretion disk by a hot coronae above the surface
\citep[][]{BisnovatyiBlinnikov1976,HaardtMaraschi1991,HaardtMaraschi1993,SvenssonZdziarski1994,Zdziarskietal1999}.
The corona, once formed, is usually thought to be located in a
compact region near the innermost stable circular orbit
\citep[e.g.][]{ReisMiller2013, DexterBlaes2013}.  However, despite the rich observational
evidence for the existence of hot coronae, there is no general
consensus regarding the mechanism by which such coronae are formed.
Many models \citep[][]{HaardtMaraschi1991,SvenssonZdziarski1994}
simply assume that some fraction of the energy liberated by
accretion is dissipated in optically thin regions.  Magnetic
reconnection is often invoked as the energy dissipation mechanism
in the corona
\citep[e.g.][]{Galeev1979,GoodmanUzdensky2008,UzdenskyGoodman2008}.  On
the other hand, magneto-rotational instability (MRI) is now understood
to be the mechanism that drives angular momentum transport and
energy dissipation in black hole accretion disks
\citep[][]{BalbusHawley1991,BalbusHawley1998}.  It is therefore of
great interest to explore whether the MRI naturally leads to high
temperature, magnetic pressure supported corona, and if so, by what mechanism.

In fact, early numerical simulations that studied the vertical
structure of accretion disks with MRI turbulence
\citep[][]{MillerStone2000} showed that indeed a strongly magnetized
region was formed in the low density upper layers of the
disk. However, these simulations adopted an isothermal equation of
state, and therefore the gas temperature in the upper layers could
only be estimated {\em ex post facto}.  Using the flux-limited
diffusion (FLD) module \citep{TurnerStone2001} in the ZEUS code,
\cite{Hiroseetal2006} studied the vertical temperature structure of
accretion disks with MRI turbulence and radiative cooling.  Although
they also found magnetic pressure supported low density regions
emerged in the upper layers of the disk, the gas temperature in these
layers was always much smaller than at the disk mid-plane.  Detailed
spectral modeling \cite{Blaesetal2006} based on the vertical profiles
of dissipation and magnetic support taken from these simulations
showed a slight hardening of the spectrum compared to previous models
that neglected these effects, but no sign of a much hotter corona.
Subsequent models computed using the same code
\citep[][]{Kroliketal2007, Blaesetal2007, Hiroseetal2009,
  Blaesetal2011} that explored a wide range of ratios between
radiation and gas pressure always found similar temperature profiles.
At the same time, a common property of these simulations is a
significant increase in dissipation rate per unit mass in the surface
regions, suggesting a temperature inversion might be possible under
the right circumstances.

All previously published radiation MHD simulations of the MRI have
considered a large surface density so that the total electron
scattering optical depth from the disk mid-plane to the surface is
$\sim 10^4$.  In this paper, we report on new radiation MHD simulations
performed using the full-transport radiation transfer module in
Athena \cite[][]{Davisetal2012, Jiangetal2012}.  We find that when the disk
surface density is decreased so that the total electron scattering
optical depth is only $\sim 10^2$, a strong temperature inversion
forms above the disk surface due to dissipation of the turbulence
in magnetically supported regions {\em which are optically thin}.
This temperature inversion is (in many respects) consistent with
the observationally inferred properties of accretion disk coronae,
and we will refer to it as a corona throughout this work.  Due to
the thermal runaways that we generically observe in radiation
pressure dominated simulations \citep[][]{Jiangetal2013c}, we focus
solely on gas pressure dominated disks in this paper.  This constraint
prevents the simulations from producing the very high temperatures ($\sim
10^9$ K) that are observationally inferred, but does not prevent
an initial exploration of this potentially new mechanism for corona
formation.

\section{Method}
We adopt the local shearing box approximation \citep[][]{HGB1995,Jiangetal2013c}, which means 
we study a local patch of the accretion disk in a frame rotating with orbital frequency $\Omega$
at a fiducial radius $r_0$ from the central BH with black hole mass $\mbh=6.62\msun$. 
Curvature of the orbit is neglected so the radial, azimuthal and vertical directions are represented by 
the local Cartesian coordinate $(x,y,z)$ with unit vector ($\bm{\hat{i}}$, $\bm{\hat{j}}$, $\bm{\hat{k}}$) 
respectively.  The vertical component of the gravitational force from the 
BH under the thin disk approximation is included as $-\rho\Omega^2z\bm{\hat{k}}$. 
Energy changes of the gas due to Compton scattering 
is approximated as  \citep[][]{Hiroseetal2009} $\Delta E_{s}=-4cE_r\rho\kappa_{es}(T-T_r)/T_e$, where $c$ is the speed of light, $E_r$ is 
the radiation energy density, $\rho$ is gas density and electron scattering opacity 
$\kappa_{es}=0.33$  cm$^2$ g$^{-1}$,  $T$ is the gas temperature, $T_r$ is the radiation temperature defined as 
$T_r\equiv\left(E_r/a_r\right)^{1/4}$ with radiation constant $a_r=7.57\times10^{15}$ erg cm$^{-3}$ K$^{-4}$. 
The equivalent electron temperature is defined as $T_e\equiv m_ec^2/k_B=5.94\times10^9\ \text{K}$, where 
$m_e$ is the electron mass and $k_B$ is the Boltzmann  constant. The radiation field is always assumed to 
be a Planck distribution in this formula for Compton scattering.

The complete set of radiation MHD evolutionary equations we solve are given by equations (2) and (3) of 
\cite{Jiangetal2013c}. 
Our solutions are computed using the Godunov radiation MHD code based on a variable Eddington tensor (VET) 
as described and tested by \cite{Jiangetal2012} and \cite{Davisetal2012}, with the improvements 
described at the Appendix of \cite{Jiangetal2013b}. The adiabatic index is chosen to be $\gamma=5/3$ with 
mean molecular weight $0.6$. Plank-mean free-free
absorption opacity $\kappa_{aP}=3.7\times10^{53}\left(\rho^9/E_g^7\right)^{1/2}$
cm$^2$ g$^{-1}$ and Rosseland-mean free-free absorption opacity
$\kappa_{aF}=1.0\times10^{52}\left(\rho^9/E_g^7\right)^{1/2}$ cm$^2$ g$^{-1}$, where 
$E_g$ is the gas internal energy density.  Unit of the magnetic field is chosen such that 
magnetic permeability is one \citep[][]{Stoneetal2008}.

\subsection{Initial and Boundary Conditions}
In order to see the effects of different surface densities on the vertical structure of accretion disks, 
we compare two simulations. Simulation A is located at 
$r_0=30\left(G\mbh/c^2\right)$ with $\Omega=190.1$ s$^{-1}$, where $G$ is the gravitational 
constant, $c$ is the speed of light. Total electron scattering optical depth from the disk mid-plane to the surface of the disk 
for this simulation is $\tau_e=288.4$, which corresponds 
to a total disk surface density $1.75\times 10^3$ g cm$^{-2}$.  We pick the initial disk mid-plane density, temperature and pressure 
to be $\rho_0=10^{-3}$ g cm$^{-3}$, $T_0=10^7$ K and $P_{g,0}=1.39\times 10^{12}$ dyn cm$^{-2}$ respectively. 
The length scale is chosen to be $H=c_{s,0}/\Omega=1.96\times10^5$ cm, 
where $c_{s,0}$ is the isothermal sound speed corresponding to $T_0$. 
Simulation B is located at $r_0=300\left(G\mbh/c^2\right)$ with $\Omega=6.0$ s$^{-1}$. Electron scattering optical 
depth for this run is $\tau_e=1.03\times 10^4$, which corresponds to $36$ times the disk surface density of simulation A. 
The initial disk mid-plane density, temperature, pressure for simulation B are $\tilde{\rho_0}=1.12\times 10^{-2}$  g cm$^{-3}$, $\tilde{T}_0=2.89\times 10^6$ K, 
$\tilde{P}_{g,0}=4.41\times 10^{12}$ dyn cm$^{-2}$ respectively. The corresponding disk scale height is then $\tilde{H}=3.53\times 10^6$ cm. 
Parameters for simulation B are chosen to match the simulations described by \cite{Hiroseetal2006} so that the results can be directly compared. 
The initial parameters for the two simulations are summarized in Table \ref{Table:parameters}.

\begin{table*}[htp]
\centering
\caption{Summary of the Simulation Parameters}
\begin{tabular}{ccccccccc}
\hline
Label 	&	$\Omega/s^{-1}$ & $\Sigma/ 10^3$ g cm$^{-2}$ 	& $\rho_0$/ $10^{-3}$ g cm$^{-3}$ 	& $T_0$/$10^7$ K  	&  $H$/$10^5$ cm 	& Box$/H$  			& Grids/$H$ & $\<P_r/P_g\>$  \\
\hline
A		& 190	& $1.75$					& 1.0				&$1.0$			&$1.96$			& $1\times8\times4$	&$64^2\times 32$	&	0.0052 \\
B		& 6.0		& $63.0$					& 11.2			&$0.29$			&$35.3$			& $1\times16\times4$	&$32^3$		&	0.25	\\
\hline
\end{tabular}
\label{Table:parameters}
\begin{tablenotes}
\item    Note: The box size and grids are for $x$, $z$ and $y$ directions respectively. The ratio $\<P_r\>/\<P_g\>$
is the time and horizontally averaged value at the disk mid-plane of each simulation. 
\end{tablenotes}
\end{table*}
Given the above mid-plane parameters, we calculate the initial vertical profile of the disk according to the hydrostatic and diffusion equations, with assumed local 
dissipation rate proportional to $\rho/\sqrt{\tau}$, where $\tau$ is the electron scattering optical depth measured from the surface of the disk \citep[][]{Jiangetal2013c}. 
For the magnetic field, we initialize two oppositely twisted flux tubes with the same net azimuthal flux. The $B_x$ and $B_z$ components of the field are generated by the
vector potential $A_B(x,y,z)=-\text{sign}(z)B_0\left[1+\cos\left(\pi
r\right) \right]/\left(32\pi\right)$ for $r\leq 0.25$, where $r\equiv
\left[x^2+\left(|z|-0.25\right)^2\right]^{0.5}$, while the $B_y$ component
is initialized from $B_y=\left(B_0^2/2-B_x^2-B_z^2\right)^{1/2}$  for
$|z|<0.8$. 
The ratio between gas pressure and magnetic pressure is $10$ initially at the disk mid-plane. 
We also adopt a density floor $5\times 10^{-6}$ times the initial disk mid-plane density throughout the numerical integration to avoid very small time step. 
The simulation box size along the $x$, $y$ and $z$ directions are $L_x=H$, $L_y=4H$ and $L_z=8H$ for simulation A and $L_x=\tilde{H}$, 
$L_y=4\tilde{H}$, $L_z=16\tilde{H}$ for simulation B. For simulation A, we use resolution 
$64$ grids per $H$ for the $x$ and $z$ directions but $32$ grids per $H$ for the $y$ direction as azimuthal structures are smoother 
due to the shearing.  For simulation B, the resolution is $32$ grids per $\tilde{H}$ for all three directions. 
  Larger vertical box size is used for simulation B in order to include the photosphere as surface density is larger for this one. 
The same resolution as simulation A is also tried for Simulation B with Eddington approximation, which gives very similar 
vertical profiles for the gas quantities.
The boundary conditions are the same as described in Section 3 of \cite{Jiangetal2013c}. For the short characteristics
module used to calculate the VET, we use $10$ angles per octant. 

\section{Results}
\label{sec:result}

During the linear growth phase of the MRI in the first $~\sim 10$ orbits of both simulations, the disks 
cool radiatively through the surfaces and slowly collapse. After the MRI saturates, the disks are heated by the 
dissipation of turbulence.  Subsequently, the disks adjust their vertical structure to reach a thermal equilibrium 
state in which radiative cooling and turbulent heating are balanced.  Since the disks are 
supported by gas pressure, they are thermally stable \cite[][]{ShakuraSunyaev1976, Jiangetal2013c}. We run 
simulation A for more than $120$ orbits with average cooling time to be $5$ orbits. Simulation 
B runs to $200$ orbits, much longer than the average cooling time of $10$ orbits. 
Here the cooling time is defined to be the ratio between the total energy density inside the 
simulation domain and the cooling rate from the surfaces of the domain.  
If we map the local shearing box simulations to global disk structures by assuming cooling is balanced 
by the release of gravitational energy, the steady state 
of simulation A roughly corresponds to a thin disk model with $\alpha=0.03$ and $3.8\times10^{-5}$ 
Eddington accretion rate, while simulation B corresponds to a thin disk model with the same $\alpha$ 
but $0.12$ Eddington accretion rate. That is why although simulation A is closer to the central black hole, 
it can have a smaller surface density and lower mid-plane temperature.

\begin{figure}[htp]
\centering
\vspace{0.6cm}
\includegraphics[width=1.0\hsize]{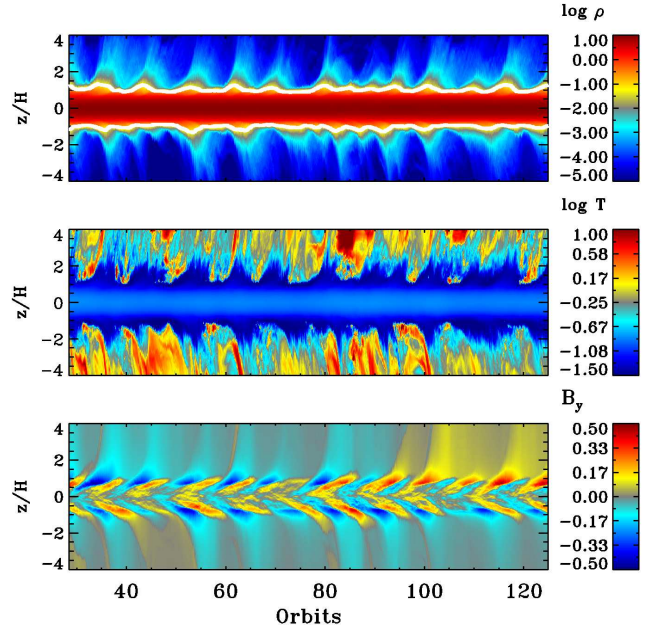}
\caption{Space-time diagram of the density $\rho$ (top panel, in unit of $\rho_0$), 
gas temperature $T$ (middle panel, in unit of $T_0$)
and azimuthal magnetic field $B_y$ (in unit of $\sqrt{2P_0}$) for the simulation that forms a corona. 
The white line in the top panel shows the location of the photosphere for electron scattering 
opacity. }
\label{STplot}
\end{figure}

\begin{figure}[htp]
\centering
\includegraphics[width=1.0\hsize]{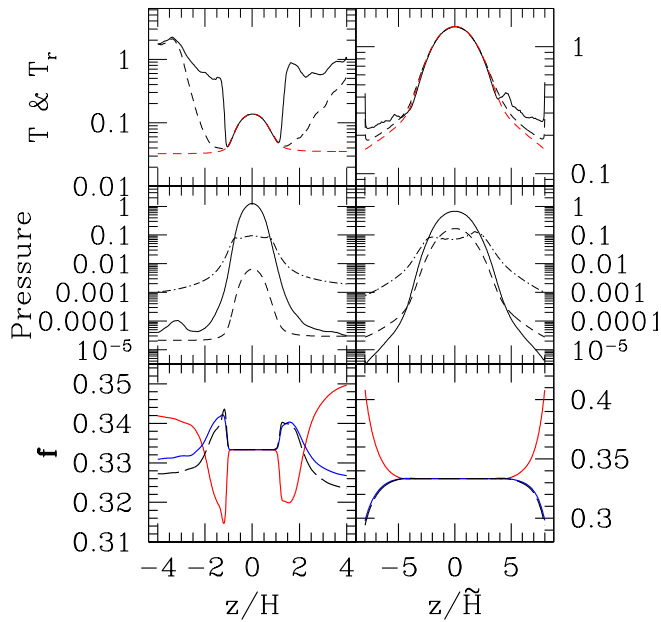}
\caption{\emph {Left:}
Horizontally and time averaged vertical profiles of the gas ($T$, black lines) and radiation ($T_r$, red lines) 
temperature (top panel), magnetic ($P_B$, dashed-dot line), radiation ($P_r$, dashed line) and gas ($P_g$, solid line) 
pressure (middle panel), three components of Eddington tensor $f_{xx}$ (black line), $f_{yy}$ (blue line) and 
$f_{zz}$ (red line) (bottom panel), for simulation A. In the top panel, the solid black line is the volume weighted average 
gas temperature while the dashed black line is the density weighted average temperature. 
The temperature unit is $T_0=10^7$ K while the pressure unit is $P_{g,0}=1.39\times10^{12}$ dyn cm$^{-2}$. 
\emph{Right:} The same as the left panels but for simulation B. The units for temperature and pressure for 
this simulation are $\tilde{T}_0=2.89\times 10^6$ K and $\tilde{P}_{g,0}=4.41\times 10^{12}$ dyn cm$^{-2}$.
}
\label{TPProfile}
\end{figure}

The evolution histories for density $\rho$, temperature $T$ and azimuthal magnetic field  $B_y$ 
for simulation A  are shown in a space-time diagram in Figure \ref{STplot}, generated  
by averaging quantities across horizontal planes and then plotting the resulting 
vertical profiles versus time. Consistent 
with previous simulations with much larger surface density \citep[][]{Hiroseetal2006,Jiangetal2013c}, 
the sign of $B_y$ flips roughly every $10$ orbits, which is the well-known butterfly diagram 
observed in almost all the vertically stratified shearing box simulations 
\citep[e.g.,][]{Stoneetal1996,MillerStone2000,Davisetal2010,Jiangetal2013c}. 
The space-time diagram for density also shows very similar 
structures. As the magnetic field rises buoyantly, it carries along some gas which is denser than the surrounding coronal
material; this gas later
falls back towards the disk mid-plane. During the whole simulation, the photosphere is well inside the simulation box, 
as labeled by the white line at the top panel of Figure \ref{STplot}. Simulation B also shows 
very similar behavior. 

The most dramatic difference between simulations A and B is the space time diagram of the gas 
temperature. Simulation B is similar to previous MRI simulations with radiation transfer  \citep[][]{Hiroseetal2006} 
in that gas temperature peaks at the disk mid-plane and drops with height. However, for simulation 
A, as shown in Figure \ref{STplot}, the gas temperature above the electron scattering 
photosphere is dramatically larger than the gas temperature at the disk mid-plane. The gas 
temperature is maximum during the periods when the magnetic field rises buoyantly. 
This correlation exists because dissipation in the corona is directly from the magnetic 
energy density, which is not amplified locally at each height but carried here from 
the disk mid-plane.

\begin{figure}[htp]
\centering
\includegraphics[width=1.0\hsize]{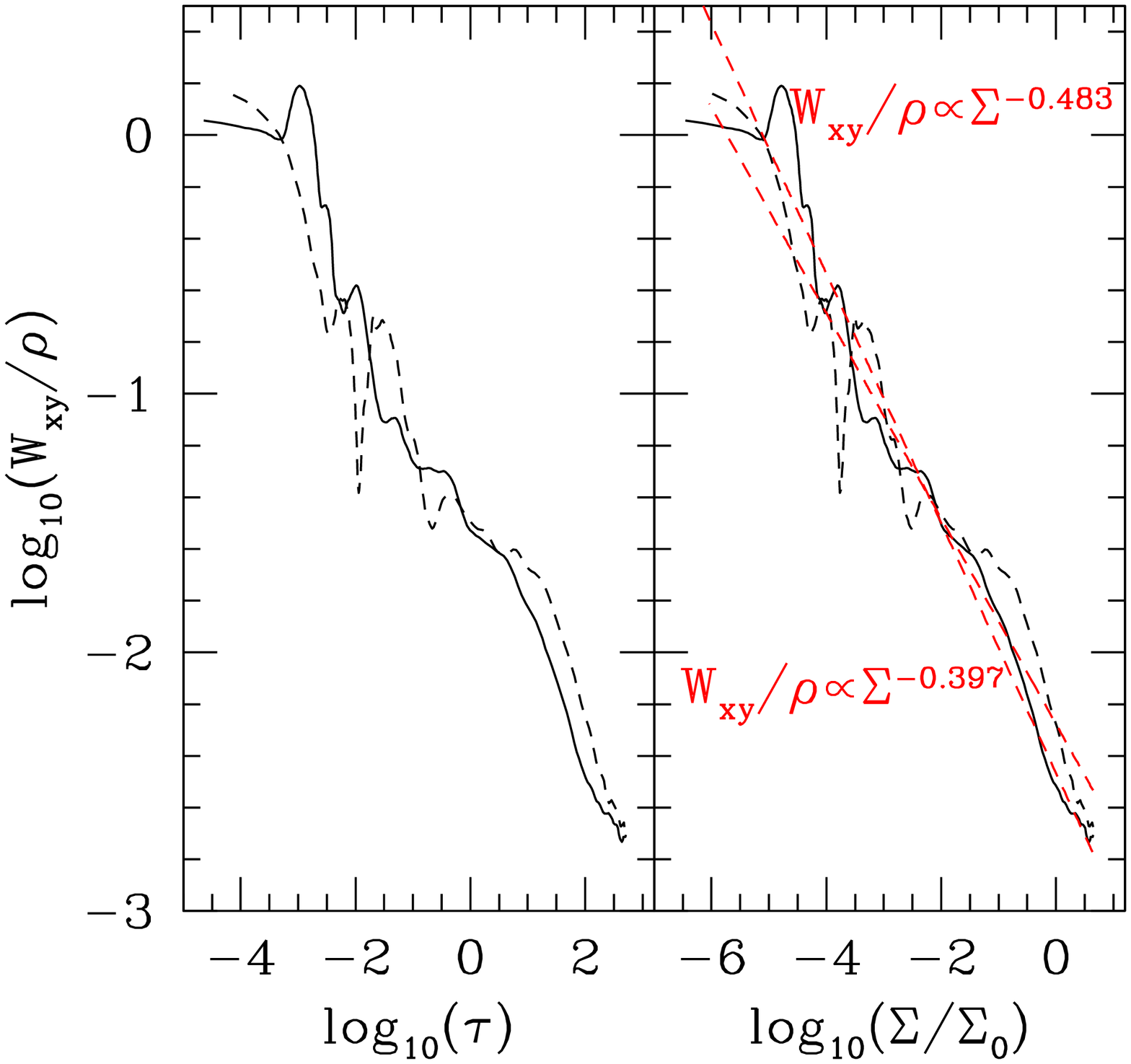} \\
\vspace{-0.9cm}
\includegraphics[width=1.0\hsize]{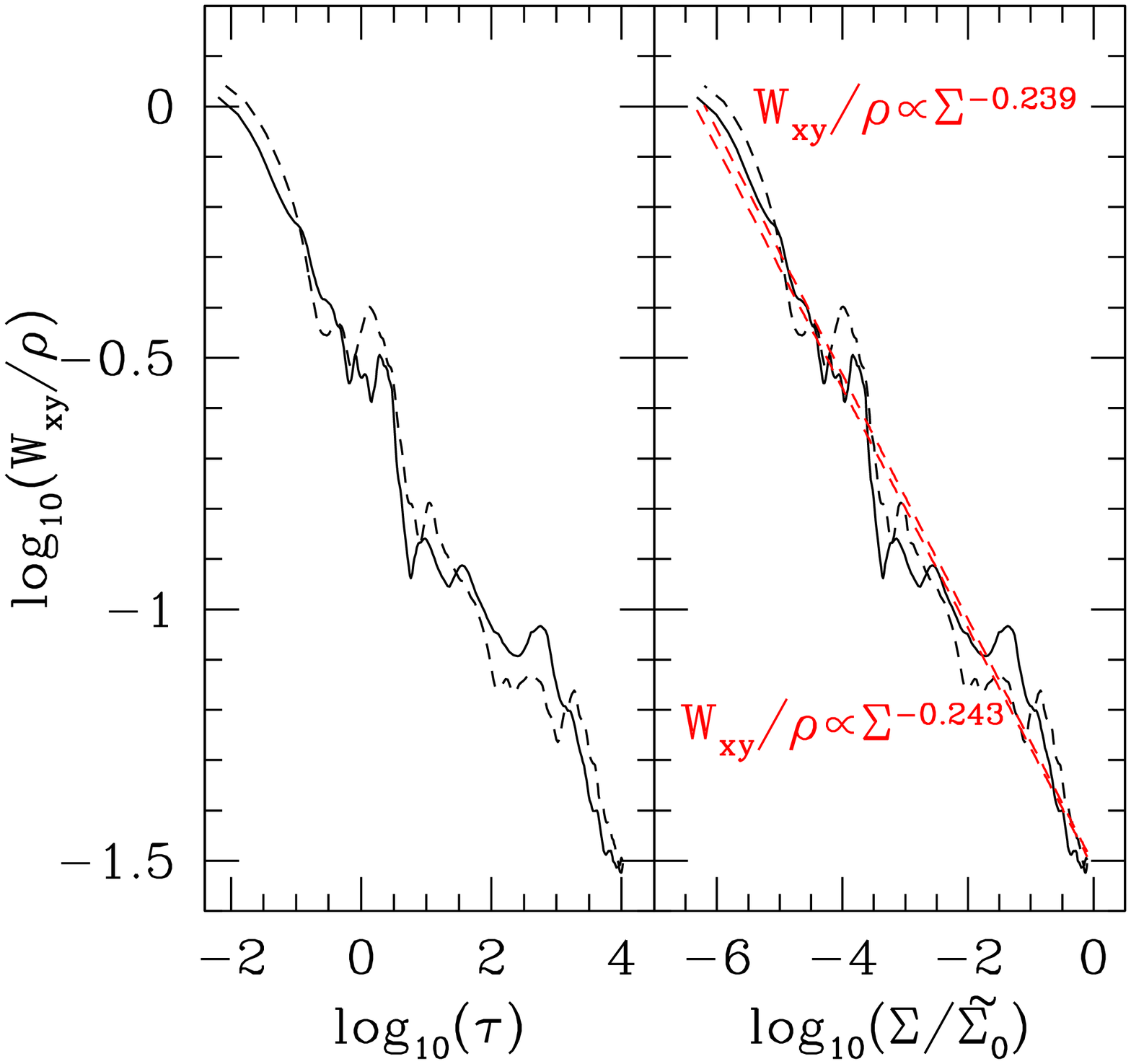} 
\caption{\emph{Top:} Profiles of the stress per unit mass produced by the MRI turbulence 
as a function of optical depth measured from the disk surface (left panels) and 
the disk surface density (right panels). 
 The black solid lines are for the part above 
the disk mid-plane while the black dashed lines are for the part below the disk 
mid-plane. The red dashed lines are best fitting relations between $W_{xy}/\rho$ 
and $\Sigma$ for the top and bottom of the disk. \emph{Bottom:} The same as the top 
two panels but for simulation B.}
\label{dissipationProfile}
\vspace{0.1cm}
\end{figure}

\subsection{Vertical Profiles of the Disk}

To investigate the vertical structure of the disk quantitatively, we time-average the vertical profiles
between $25$ and $120$ orbits for simulation A and $50$ to $200$ orbits for simulation B. 
The resulting time-averaged vertical profiles of gas ($T$) and radiation ($T_r$) temperature, 
gas ($P_g$), radiation ($P_r$) and magnetic pressure ($P_m$), as well as three components of the Eddington tensor 
($f_{xx}$, $f_{yy}$, $f_{zz}$) are shown in Figure 
\ref{TPProfile}. Within the disk photosphere, gas and radiation temperature are coupled 
and they both decrease with height from the disk mid-plane. In this optically thick region, the Eddington tensor is 
close to $1/3\bI$ as expected and $P_g$ is much larger than $P_m$ and $P_r$. 
Once above the photosphere, $T$ and $T_r$ decouple. 
For simulation B, $T$ continues to decrease with height. The rapid change near the 
boundary is likely caused by the boundary condition.  The whole temperature profile is 
consistent with Figure 3 of \cite{Hiroseetal2006}. However, for simulation A, 
radiation temperature $T_r$ stays almost flat above the photosphere
 while gas temperature increases very quickly 
by a factor of $\sim 30$. Gas pressure also drops quicker than magnetic pressure above the photosphere and this 
region is supported primarily by magnetic pressure. High gas temperature with strong magnetic pressure support are 
the defining  characteristics of the corona observed in our simulations. 
The sharp increase of the gas temperature also causes peaks in the horizontal components of the Eddington tensor and a dip in the vertical component. 
This is because the rising temperature in combination with the falling density produces a localized (in $z$) maximum in the emissivity.  
At nearly horizontal viewing angles, one has longer line-of-sight through this emissivity ``bump" yielding larger intensities for horizontal rays and lower intensities for vertical rays. 
This projection effect becomes less important closer to the top of the domain and the intensity becomes more nearly isotropic and eventually slightly limb-darkened.
After the gas temperature becomes flat, $f_{zz}$ 
becomes larger than $1/3$ while $f_{xx}$ and $f_{yy}$ becomes smaller than $1/3$, which are consistent with 
profiles of Eddington tensor in simulation B.

\begin{figure}[htp]
\centering
\includegraphics[width=1.0\hsize]{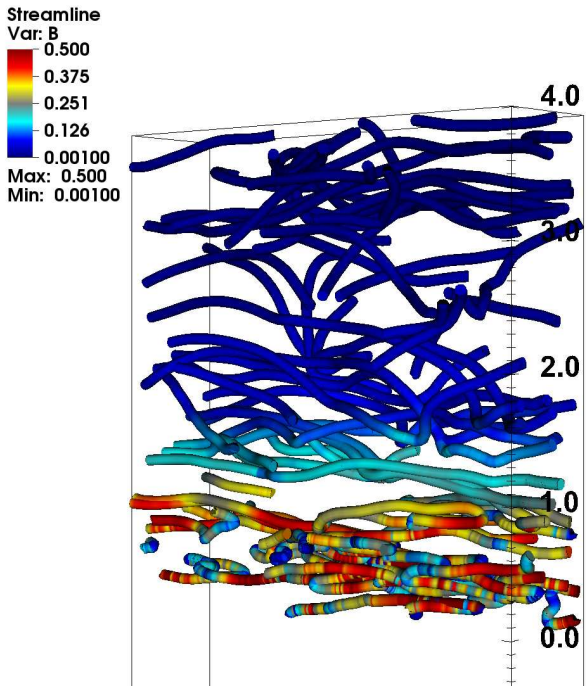} 
\caption{Magnetic field streamline at $95$ orbits of the simulation with corona. Color represents magnitude of the 
magnetic field.  }
\label{streamline}
\vspace{0.1cm}
\end{figure}

\begin{figure}[htp]
\centering
\includegraphics[width=1.0\hsize]{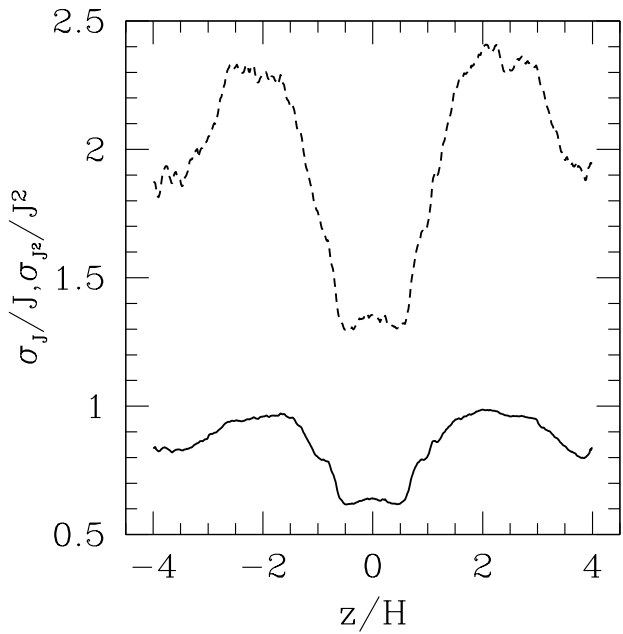} 
\caption{Time averaged vertical profiles of 
$\sigma_J/J$ (solid line) and $\sigma_{J^2}/J^2$ (dashed line) for simulation A, where 
$J$ and $J^2$ are averaged along the horizontal plane. }
\label{current}
\vspace{0.1cm}
\end{figure}

\subsection{The Dissipation Profile}
The reason that gas temperature increases above the photosphere is that there is still significant dissipation in this region. 
Because it is optically thin, the gas cannot cool easily and its temperature rises. Because of the different temperature 
profiles between simulation A and B, we naturally expect different dissipation profiles for the two simulations with 
significantly different surface densities. For local shearing box simulations of MRI, the energy input is proportional 
to the sum of Maxwell and Reynolds stress $W_{xy}$ from MRI turbulence \citep[][]{HGB1995,Jiangetal2013b}.   The vertical
profiles of dissipation can therefore be directly measured
from our simulations, removing one of the largest  
uncertainties inherent in emission models of accretion disks \citep[][]{Davisetal2005,TaoBlaes2013}.
The horizontally and time averaged profiles of 
stress \emph{per unit mass} as a function of total optical depth $\tau$ and surface density $\sigma$ for simulation A and B are shown 
in Figure \ref{dissipationProfile}. As this figure shows, $W_{xy}/\rho$ increases rapidly with height, while the standard 
$\alpha$ disk model assumes it is a constant. By comparing simulation A and B, it is clear that the vertical profile of 
$W_{xy}/\rho$ becomes steeper as surface density decreases. At the same time, the position of the photosphere 
moves closer to the disk mid-plane. In simulation B, only $0.0094\%$ of the total mass is located above the 
photosphere, which only contains $0.085\%$ of the energy dissipation. However for simulation A with much smaller surface density, 
the optically thin photosphere includes  $0.26\%$ of the total mass and $3.4\%$ of the total dissipation. 
Clearly, a larger fraction of energy is dissipated in the optically thin region in simulation A, and ultimately
this is why a hot corona is formed. 
This also suggests that the typical scale height of dissipation does not scale with density scale height. This is not 
surprising as dissipation in MRI is ultimately related to the magnetic field, which always tends to rise up to the low density region 
due to buoyancy.

To provide some context for how dissipation occurs in the simulation, Figure \ref{streamline} 
shows a snapshot of the typical magnetic field structure in the top half of the box. 
The magnetic field is mainly along the toriodal direction 
in the optically thick mid-plane, where the Maxwell stress is 
the largest. In the magnetic pressure supported optically thin photosphere, a significant vertical component of the magnetic field 
can be formed due to Parker instability \citep[][]{Blaesetal2007,HiroseTurner2011}. To investigate the possible association of these structure with
dissipation, we plot the time averaged vertical profiles of the current density $J$, $J^2$, and their standard deviations normalized to horizontally averaged values  
in Figure \ref{current}.  We expect that regions where the fluctuations in these quantities are large are likely to be dominated by dissipation at small scales.  
The figure demonstrates the  horizontal fluctuations are larger in the corona region compared with the disk mid-plane for both $J$ and $J^2$, 
which suggests that small scale dissipation is still dominant in the corona.  We caution, however, that  it is not clear if fluctuations in $J$ or $J^2$ are indeed a good indicator of the scales in which dissipation occurs, and
moreover this dissipation may not necessarily be related to reconnection.  A more quantitative exploration of the role of reconnection in MRI
turbulence in both the mid-plane and corona is beyond the scope of this work.

\section{Discussions and Conclusions}
By comparing two different simulations with MRI turbulence and radiation transfer, we first 
confirm that when the surface density is so large that the total electron 
scattering optical depth is $\sim 10^4$, there is very little dissipation in the optically 
thin part and the temperature in the magnetic pressure supported region is always much smaller than 
the disk mid-plane. When the surface density of the disk is small enough such that the total electron 
scattering optical depth is only $\sim 200$, a non-negligible fraction of dissipation from MRI occurs above the 
photosphere, and a high temperature, magnetic pressure supported corona is self-consistently generated. 
Note that the disk is still in the gas pressure dominated regime and the temperature 
in the corona is only $\sim 10^7$ K in this particular run. Since this is still much smaller than what is required 
to make the observed hard X-rays, we do not claim to have reproduced the corona of observed systems. 
Observationally consistent coronal models for most observed systems will require a larger
fraction of the dissipation above the photosphere than found in simulation A ($\sim 3.4\%$). Nevertheless,
we consider this a useful demonstration that MRI turbulence can produce
temperature inversions in magnetically dominated surface regions in simulations with realistic thermodynamics. 
This also motivates future studies to see how the structures of the accretion disks will be changed if more and 
more energy is dissipated in the optically thin part when the disk surface density decreases further.



Figure \ref{TPProfile} shows a notable contrast between the volume average and density-weighted average of the temperature
in the surface regions.  Near the surface, there is a significant anti-correlation between the temperature and density that 
leads to the volume average temperature being higher than the density-weighted average.  Dissipation in this region is dominated
by work done by magnetic stresses and numerical reconnection.  The associated heating is rather inhomegenous  as indicated by Figure \ref{current}
and tends to deposit
a larger amount of energy per unit mass in lower density regions, giving rise to the temperature-density anti-correlation.  Such `patchy' heating
is interesting because it may be necessary to explain the hard X-ray tails observed in high/soft or very high state X-ray binaries
\citep[see e.g. sec. 5.3.1 of][]{Doneetal2007}.  In these systems, the hard
spectral slopes are suggestive of high temperatures (or high electron energies in a non-thermal electron distribution), but
the fraction of the power emitted in the component suggests that only a small fraction of photospheric photons are scattered.
Such a geometry arises naturally if a significant fraction of the heating occurs in a small fraction of the volume, as the simulations
imply.

One limitation of these shearing box simulations is that they consider only a small patch of the disk, so the global structure of the 
accretion disk and corona cannot be studied.  Post processing of global MHD simulations with Monte Carlo radiative transfer calculations 
suggests that many aspects of the observed hard X-ray spectrum can be reproduced \citep[][]{Schnittmanetal2013}.  One drawback of these
calculations is that they rely on heuristic cooling function to maintain a thin disk in the simulations \citep{Nobleetal2009}.  In the future, 
we hope to perform a similar analysis with global radiation MHD simulations, particularly since many of the observed systems are
in the radiation pressure dominated regime.  Here, we simply discuss how these local patches of the disk would fit into standard
accretion disk models.

In the $\alpha$ disk model, the surface density of the disk decreases
as the disk effective and mid-plane temperatures decrease in the gas
pressure dominated regime \citep[][]{ShakuraSunyaev1973}, consistent
with both ZEUS and Athena simulation results
\citep{Hiroseetal2009b,Jiangetal2013c}.  In a global accretion disk
model, this corresponds to reducing the accretion rate and total
luminosity. But at the same time, we have shown that a larger fraction
of energy is dissipated in the corona region. Therefore, the ratio
between luminosity from high energy band and low energy band, or the
spectral index, would increase when the total luminosity decreases.
The transition from the high soft state to the low-hard state in X-ray
binaries is a case where spectrum hardens in this manner as the
luminosity decreases \citep[][]{RemillardMcClintock2006}.  Even if the
inner disk is radiation dominated at the beginning of the transition,
it should become gas pressure dominated as the accretion rate drops so
the surface density must decrease as well.  There is some
observational evidence that during this transition, the optically
thick disk is truncated at radii larger than the last stable circular
orbit \citep[][and references therein]{Doneetal2010}.  In the inner
disk, the geometrically thin and optically thick disk must become a
geometrically thick and optically thin ADAF-like flow
\citep{NarayanYi94}. If this interpretation is correct, there is
likely to be some transition region with total electron scattering
optical depth smaller than $\sim 100$, which corresponds to the
parameters adopted in our simulations. 

For the radiation pressure dominated regime, the $\alpha$ disk model
predicts that the disk effective and midplane temperatures will
increase as the disk surface density decreases. However, our
shearing box simulations of the radiation pressure dominated regime
show thermal runaways \citep[][]{Jiangetal2013c} and it remains
unclear how these runaways might saturate in a global disk.
Nevertheless, \cite{Jiangetal2013c}, found that in the radiation
pressure dominated regime, the dissipation scale height changes slower
than the change of density scale height. Therefore a larger fraction
of energy will be dissipated at high altitude when disk surface
density decreases. On one hand, this may be inconsistent with some
high/soft state observations which have relatively low coronal
contribution. On the other hand, the fact that the very high state (or
steep power law state) has a strong tail and only seems to occur at
high luminosities \citep[][]{RemillardMcClintock2006}, suggests that
the surface density of the accretion disk is only small enough to
enable the formation of corona when the accretion rate is very large
in the radiation pressure dominated regime. 
Therefore, it will still be interesting to see in the radiation pressure 
dominated case before the thermal runaway completely changes 
the structures of the disk, how corona will be formed 
when the surface density is decreased. This will be studied 
with future shearing box simulations.

Most AGN are expected to be in the radiation pressure dominated
regime, except at low Eddington rates where a transition to an
ADAF-like flow occurs.  Therefore, we can only speculate on the
observational implications of our current, gas pressure dominated
results.  We note that X-ray and optical observations of AGNs do
indicate a trend of hardening of the spectral index as luminosity
decreases \citep[][]{Steffenetal2006,Justetal2007}, although the black
hole mass is uncertain so it is unclear what the relevant Eddington
ratios are in these systems. Recently, a new population of quasars
have been found to show much weaker X-ray emission compared with other
AGNs with similar luminosity range
\citep[][]{Wuetal2011,Luoetal2013}. Observationally, it is still
debated whether these quasars are intrinsically X-ray weak or
obscured. If the AGN X-ray emission is indeed from the corona
generated by MRI turbulence as found here, then these quasars may be
intrinsic X-ray weak when most of the dissipation happens inside the
photosphere and the corona becomes very weak or disappears.  Finally,
we note that the increased dissipation per unit mass near the disk
surface may be the mechanism needed to generate the continuum soft
X-ray excess inferred in some AGN \citep{Czernyetal03,Doneetal2012}.
Ultimately, testing these hypotheses will require global simulations
of accretion flows in both the gas and radiation pressure dominated
regimes.



\section*{Acknowledgements}
Y.F.J thanks Jenny Greene, Yue Shen and Jianfeng Wu for helpful discussions 
on the X-ray and optical observations of AGNs, and J. Goodman for discussions which
motivated this project. 
 We also thank the anonymous referee for helpful comments that improved the paper. 
This work was supported by the
NASA ATP program through grant NNX11AF49G, and by computational resources
provided by the Princeton Institute for Computational Science and Engineering. 
Some of the simulations were performed on the Pleiades Supercomputer 
provided by NASA. 
Y.F.J. is supported by NASA through 
Einstein Postdoctoral Fellowship grant number 
PF-140109 awarded by the Chandra X-ray Center, 
which is operated by the Smithsonian Astrophysical 
Observatory for NASA under contract NAS8-03060. 
This work was also supported in part by the U.S. National Science
Foundation, grant NSF-OCI-108849 and NSF-AST-1333091.
 
 


\bibliographystyle{astroads}
\bibliography{Corona}

\begin{thebibliography}{48}
\expandafter\ifx\csname natexlab\endcsname\relax\def\natexlab#1{#1}\fi
\expandafter\ifx\csname href\endcsname\relax
  \def\href#1#2{}\fi
\expandafter\ifx\csname urllinklabel\endcsname\relax
  \def\urllinklabel{[LINK]}\fi
\expandafter\ifx\csname adsurllinklabel\endcsname\relax
  \def\adsurllinklabel{[ADS]}\fi

\bibitem[{{Balbus} \& {Hawley}(1991)}]{BalbusHawley1991}
{Balbus}, S.~A. \& {Hawley}, J.~F. 1991, \apj, 376, 214


\bibitem[{{Balbus} \& {Hawley}(1998)}]{BalbusHawley1998}
---. 1998, Reviews of Modern Physics, 70, 1


\bibitem[{{Bisnovatyi-Kogan} \& {Blinnikov}(1976)}]{BisnovatyiBlinnikov1976}
{Bisnovatyi-Kogan}, G.~S. \& {Blinnikov}, S.~I. 1976, Soviet Astronomy Letters,
  2, 191


\bibitem[{{Blaes} {et~al.}(2007){Blaes}, {Hirose}, \& {Krolik}}]{Blaesetal2007}
{Blaes}, O., {Hirose}, S., \& {Krolik}, J.~H. 2007, \apj, 664, 1057


\bibitem[{{Blaes} {et~al.}(2011){Blaes}, {Krolik}, {Hirose}, \&
  {Shabaltas}}]{Blaesetal2011}
{Blaes}, O., {Krolik}, J.~H., {Hirose}, S., \& {Shabaltas}, N. 2011, \apj, 733,
  110


\bibitem[{{Blaes} {et~al.}(2006){Blaes}, {Davis}, {Hirose}, {Krolik}, \&
  {Stone}}]{Blaesetal2006}
{Blaes}, O.~M., {Davis}, S.~W., {Hirose}, S., {Krolik}, J.~H., \& {Stone},
  J.~M. 2006, \apj, 645, 1402


\bibitem[{{Czerny} {et~al.}(2003){Czerny}, {Niko{\l}ajuk},
  {R{\'o}{\.z}a{\'n}ska}, {Dumont}, {Loska}, \& {Zycki}}]{Czernyetal03}
{Czerny}, B., {Niko{\l}ajuk}, M., {R{\'o}{\.z}a{\'n}ska}, A., {Dumont}, A.-M.,
  {Loska}, Z., \& {Zycki}, P.~T. 2003, \aap, 412, 317


\bibitem[{{Davis} {et~al.}(2005){Davis}, {Blaes}, {Hubeny}, \&
  {Turner}}]{Davisetal2005}
{Davis}, S.~W., {Blaes}, O.~M., {Hubeny}, I., \& {Turner}, N.~J. 2005, \apj,
  621, 372


\bibitem[{{Davis} {et~al.}(2012){Davis}, {Stone}, \& {Jiang}}]{Davisetal2012}
{Davis}, S.~W., {Stone}, J.~M., \& {Jiang}, Y.-F. 2012, \apjs, 199, 9


\bibitem[{{Davis} {et~al.}(2010){Davis}, {Stone}, \& {Pessah}}]{Davisetal2010}
{Davis}, S.~W., {Stone}, J.~M., \& {Pessah}, M.~E. 2010, \apj, 713, 52


\bibitem[{{Dexter} \& {Blaes}(2013)}]{DexterBlaes2013}
{Dexter}, J. \& {Blaes}, O. 2013, ArXiv e-prints


\bibitem[{{Done}(2010)}]{Doneetal2010}
{Done}, C. 2010, ArXiv:1008.2287


\bibitem[{{Done} {et~al.}(2012){Done}, {Davis}, {Jin}, {Blaes}, \&
  {Ward}}]{Doneetal2012}
{Done}, C., {Davis}, S.~W., {Jin}, C., {Blaes}, O., \& {Ward}, M. 2012, \mnras,
  420, 1848


\bibitem[{{Done} {et~al.}(2007){Done}, {Gierli{\'n}ski}, \&
  {Kubota}}]{Doneetal2007}
{Done}, C., {Gierli{\'n}ski}, M., \& {Kubota}, A. 2007, \aapr, 15, 1


\bibitem[{{Elvis} {et~al.}(1978){Elvis}, {Maccacaro}, {Wilson}, {Ward},
  {Penston}, {Fosbury}, \& {Perola}}]{Elvisetal1978}
{Elvis}, M., {Maccacaro}, T., {Wilson}, A.~S., {Ward}, M.~J., {Penston}, M.~V.,
  {Fosbury}, R.~A.~E., \& {Perola}, G.~C. 1978, \mnras, 183, 129


\bibitem[{{Galeev} {et~al.}(1979){Galeev}, {Rosner}, \& {Vaiana}}]{Galeev1979}
{Galeev}, A.~A., {Rosner}, R., \& {Vaiana}, G.~S. 1979, \apj, 229, 318


\bibitem[{{Goodman} \& {Uzdensky}(2008)}]{GoodmanUzdensky2008}
{Goodman}, J. \& {Uzdensky}, D. 2008, \apj, 688, 555


\bibitem[{{Haardt} \& {Maraschi}(1991)}]{HaardtMaraschi1991}
{Haardt}, F. \& {Maraschi}, L. 1991, \apjl, 380, L51


\bibitem[{{Haardt} \& {Maraschi}(1993)}]{HaardtMaraschi1993}
---. 1993, \apj, 413, 507


\bibitem[{{Hawley} {et~al.}(1995){Hawley}, {Gammie}, \& {Balbus}}]{HGB1995}
{Hawley}, J.~F., {Gammie}, C.~F., \& {Balbus}, S.~A. 1995, \apj, 440, 742


\bibitem[{{Hirose} {et~al.}(2009{\natexlab{a}}){Hirose}, {Blaes}, \&
  {Krolik}}]{Hiroseetal2009b}
{Hirose}, S., {Blaes}, O., \& {Krolik}, J.~H. 2009{\natexlab{a}}, \apj, 704,
  781


\bibitem[{{Hirose} {et~al.}(2009{\natexlab{b}}){Hirose}, {Krolik}, \&
  {Blaes}}]{Hiroseetal2009}
{Hirose}, S., {Krolik}, J.~H., \& {Blaes}, O. 2009{\natexlab{b}}, \apj, 691, 16


\bibitem[{{Hirose} {et~al.}(2006){Hirose}, {Krolik}, \&
  {Stone}}]{Hiroseetal2006}
{Hirose}, S., {Krolik}, J.~H., \& {Stone}, J.~M. 2006, \apj, 640, 901


\bibitem[{{Hirose} \& {Turner}(2011)}]{HiroseTurner2011}
{Hirose}, S. \& {Turner}, N.~J. 2011, \apjl, 732, L30


\bibitem[{{Jiang} {et~al.}(2012){Jiang}, {Stone}, \& {Davis}}]{Jiangetal2012}
{Jiang}, Y.-F., {Stone}, J.~M., \& {Davis}, S.~W. 2012, \apjs, 199, 14


\bibitem[{{Jiang} {et~al.}(2013{\natexlab{a}}){Jiang}, {Stone}, \&
  {Davis}}]{Jiangetal2013c}
---. 2013{\natexlab{a}}, \apj, 778, 65


\bibitem[{{Jiang} {et~al.}(2013{\natexlab{b}}){Jiang}, {Stone}, \&
  {Davis}}]{Jiangetal2013b}
---. 2013{\natexlab{b}}, \apj, 767, 148


\bibitem[{{Just} {et~al.}(2007){Just}, {Brandt}, {Shemmer}, {Steffen},
  {Schneider}, {Chartas}, \& {Garmire}}]{Justetal2007}
{Just}, D.~W., {Brandt}, W.~N., {Shemmer}, O., {Steffen}, A.~T., {Schneider},
  D.~P., {Chartas}, G., \& {Garmire}, G.~P. 2007, \apj, 665, 1004


\bibitem[{{Krolik}(1999)}]{Krolik1999}
{Krolik}, J.~H. 1999, {Active galactic nuclei : from the central black hole to
  the galactic environment}


\bibitem[{{Krolik} {et~al.}(2007){Krolik}, {Hirose}, \&
  {Blaes}}]{Kroliketal2007}
{Krolik}, J.~H., {Hirose}, S., \& {Blaes}, O. 2007, \apj, 664, 1045


\bibitem[{{Luo} {et~al.}(2013){Luo}, {Brandt}, {Alexander}, {Harrison},
  {Stern}, {Bauer}, {Boggs}, {Christensen}, {Comastri}, {Craig}, {Fabian},
  {Farrah}, {Fiore}, {Fuerst}, {Grefenstette}, {Hailey}, {Hickox}, {Madsen},
  {Matt}, {Ogle}, {Risaliti}, {Saez}, {Teng}, {Walton}, \&
  {Zhang}}]{Luoetal2013}
{Luo}, B., {Brandt}, W.~N., {Alexander}, D.~M., {Harrison}, F.~A., {Stern}, D.,
  {Bauer}, F.~E., {Boggs}, S.~E., {Christensen}, F.~E., {Comastri}, A.,
  {Craig}, W.~W., {Fabian}, A.~C., {Farrah}, D., {Fiore}, F., {Fuerst}, F.,
  {Grefenstette}, B.~W., {Hailey}, C.~J., {Hickox}, R., {Madsen}, K.~K.,
  {Matt}, G., {Ogle}, P., {Risaliti}, G., {Saez}, C., {Teng}, S.~H., {Walton},
  D.~J., \& {Zhang}, W.~W. 2013, \apj, 772, 153


\bibitem[{{Miller} \& {Stone}(2000)}]{MillerStone2000}
{Miller}, K.~A. \& {Stone}, J.~M. 2000, \apj, 534, 398


\bibitem[{{Narayan} \& {Yi}(1994)}]{NarayanYi94}
{Narayan}, R. \& {Yi}, I. 1994, \apjl, 428, L13


\bibitem[{{Noble} {et~al.}(2009){Noble}, {Krolik}, \& {Hawley}}]{Nobleetal2009}
{Noble}, S.~C., {Krolik}, J.~H., \& {Hawley}, J.~F. 2009, \apj, 692, 411


\bibitem[{{Reis} \& {Miller}(2013)}]{ReisMiller2013}
{Reis}, R.~C. \& {Miller}, J.~M. 2013, \apjl, 769, L7


\bibitem[{{Remillard} \& {McClintock}(2006)}]{RemillardMcClintock2006}
{Remillard}, R.~A. \& {McClintock}, J.~E. 2006, \araa, 44, 49


\bibitem[{{Schnittman} {et~al.}(2013){Schnittman}, {Krolik}, \&
  {Noble}}]{Schnittmanetal2013}
{Schnittman}, J.~D., {Krolik}, J.~H., \& {Noble}, S.~C. 2013, \apj, 769, 156


\bibitem[{{Shakura} \& {Sunyaev}(1973)}]{ShakuraSunyaev1973}
{Shakura}, N.~I. \& {Sunyaev}, R.~A. 1973, \aap, 24, 337


\bibitem[{{Shakura} \& {Sunyaev}(1976)}]{ShakuraSunyaev1976}
---. 1976, \mnras, 175, 613


\bibitem[{{Steffen} {et~al.}(2006){Steffen}, {Strateva}, {Brandt}, {Alexander},
  {Koekemoer}, {Lehmer}, {Schneider}, \& {Vignali}}]{Steffenetal2006}
{Steffen}, A.~T., {Strateva}, I., {Brandt}, W.~N., {Alexander}, D.~M.,
  {Koekemoer}, A.~M., {Lehmer}, B.~D., {Schneider}, D.~P., \& {Vignali}, C.
  2006, \aj, 131, 2826


\bibitem[{{Stone} {et~al.}(2008){Stone}, {Gardiner}, {Teuben}, {Hawley}, \&
  {Simon}}]{Stoneetal2008}
{Stone}, J.~M., {Gardiner}, T.~A., {Teuben}, P., {Hawley}, J.~F., \& {Simon},
  J.~B. 2008, \apjs, 178, 137


\bibitem[{{Stone} {et~al.}(1996){Stone}, {Hawley}, {Gammie}, \&
  {Balbus}}]{Stoneetal1996}
{Stone}, J.~M., {Hawley}, J.~F., {Gammie}, C.~F., \& {Balbus}, S.~A. 1996,
  \apj, 463, 656


\bibitem[{{Svensson} \& {Zdziarski}(1994)}]{SvenssonZdziarski1994}
{Svensson}, R. \& {Zdziarski}, A.~A. 1994, \apj, 436, 599


\bibitem[{{Tao} \& {Blaes}(2013)}]{TaoBlaes2013}
{Tao}, T. \& {Blaes}, O. 2013, \apj, 770, 55


\bibitem[{{Turner} \& {Stone}(2001)}]{TurnerStone2001}
{Turner}, N.~J. \& {Stone}, J.~M. 2001, \apjs, 135, 95


\bibitem[{{Uzdensky} \& {Goodman}(2008)}]{UzdenskyGoodman2008}
{Uzdensky}, D.~A. \& {Goodman}, J. 2008, \apj, 682, 608


\bibitem[{{Wu} {et~al.}(2011){Wu}, {Brandt}, {Hall}, {Gibson}, {Richards},
  {Schneider}, {Shemmer}, {Just}, \& {Schmidt}}]{Wuetal2011}
{Wu}, J., {Brandt}, W.~N., {Hall}, P.~B., {Gibson}, R.~R., {Richards}, G.~T.,
  {Schneider}, D.~P., {Shemmer}, O., {Just}, D.~W., \& {Schmidt}, S.~J. 2011,
  \apj, 736, 28


\bibitem[{{Zdziarski} {et~al.}(1999){Zdziarski}, {Lubi{\'n}ski}, \&
  {Smith}}]{Zdziarskietal1999}
{Zdziarski}, A.~A., {Lubi{\'n}ski}, P., \& {Smith}, D.~A. 1999, \mnras, 303,
  L11


\end{thebibliography}

\end{CJK*}

\end{document}